# Photometric monitoring (1987 to 1994) of the gravitational lens candidate UM 425 *

F. Courbin[1,2,3], P. Magain[1]**, M. Remy[1], A. Smette[1,4], J.F. Claeskens[1,5], O. Hainaut[1,2]***, D. Hutsemékers[1], G. Meylan[2], E. Van Drom[1][†]

[1] Institut d'Astrophysique, Université de Liège, Avenue de Cointe 5, B-4000 Liège, Belgium
[2] European Southern Observatory, Karl-Schwarzschild-Straβe 2, D-85748 Garching bei München, Germany
[3] Institut d'Astrophysique de Paris, 98bis Boulevard Arago, F-75014 Paris, France
[4] Now at: Kapteyn Astronomical Institute, University of Groningen, P.O. Box 800, NL-9700 AV Groningen, The Netherlands
[5] European Southern Observatory, Casilla 19001, Santiago 19, Chile

**Abstract.** We present the results of a 7 year long photometric monitoring of two components (A & B) of UM 425, thought to be images, separated by 6.5″, of the same $z = 1.47$ quasar. These components have been imaged through an R filter in order to obtain their light curves. The photometry was obtained by simultaneously fitting a stellar two-dimensional profile on each component. The brightest image (component A, $m_R = 15.7$) shows a slow and smooth increase in brightness of 0.2 magnitude in seven years, while the faintest one (component B, $m_R = 20.1$) displays an outburst of 0.4 magnitude which lasts approximately two years. The variation of component B may be interpreted in two ways, assuming UM 425 is gravitationally lensed. If it is due to an intrinsic variation of the quasar, we derive a lower limit of 3 years on the time delay from the fact that it is not observed in component A. On the other hand, if it is a microlensing "High Amplification Event", we estimate the size of the source to be $\sim 10^{-3}$pc, in agreement with standard models of AGNs. These observations are consistent with the gravitational lens interpretation of the object.
Furthermore, all the CCD frames obtained under the best seeing conditions have been co-added, in an attempt to detect the deflector. The final R image reveals a rich field of faint galaxies in the magnitude range $m_R \sim 22 - 24$. No obvious deflector, nor any system of arcs or arclets is detected, down to a limiting magnitude of $m_R \sim 24$.

**Key words:** quasars: UM 425 — cosmology: gravitational lensing — observations

## 1. Introduction

Since the discovery of the first gravitationally lensed quasar, Q 0957+561 (Walsh, Carswell & Weymann 1979), 14 confirmed or highly suspected systems have been found (e.g. Surdej & Soucail 1993). Their photometric monitoring may yield important results, both for the study of the physical properties of quasars and for the determination of cosmological parameters.

For example, on one hand, a variation of the intrinsic luminosity of the source will be seen at different epochs in the different images. The measurement of this time delay, together with the knowledge of the system's geometry, the redshifts of the source and the deflector, and the mass distribution in the deflector, can be used to determine the Hubble constant $H_0$ (Refsdal 1964a,b).

On the other hand, brightness variations of one of the lensed images may be caused by a microlensing phenomenon. This occurs when a compact object (e.g. a star in the lensing galaxy) crosses the light path to one of the images (Chang & Refsdal 1979). The most characteristic feature of this microlensing is the so-called "High Amplification Event", showing up as a peak in the light curve of the image affected. The width of the peak corresponds roughly to the time needed for the star to cross the light beam. It can thus be used to estimate the angular size of the source (e.g. Rauch 1993).

The present paper describes the results of a photometric monitoring campaign carried out for the gravitational lens candidate UM 425 which is a system showing two images with the same redshift z=1.47, separated by 6.5″ on the sky (Meylan & Djorgovski 1989). Despite the large angular separation, and although no lensing galaxy has been detected yet, the similarity of the spectra makes this system a good gravitational lens candidate.

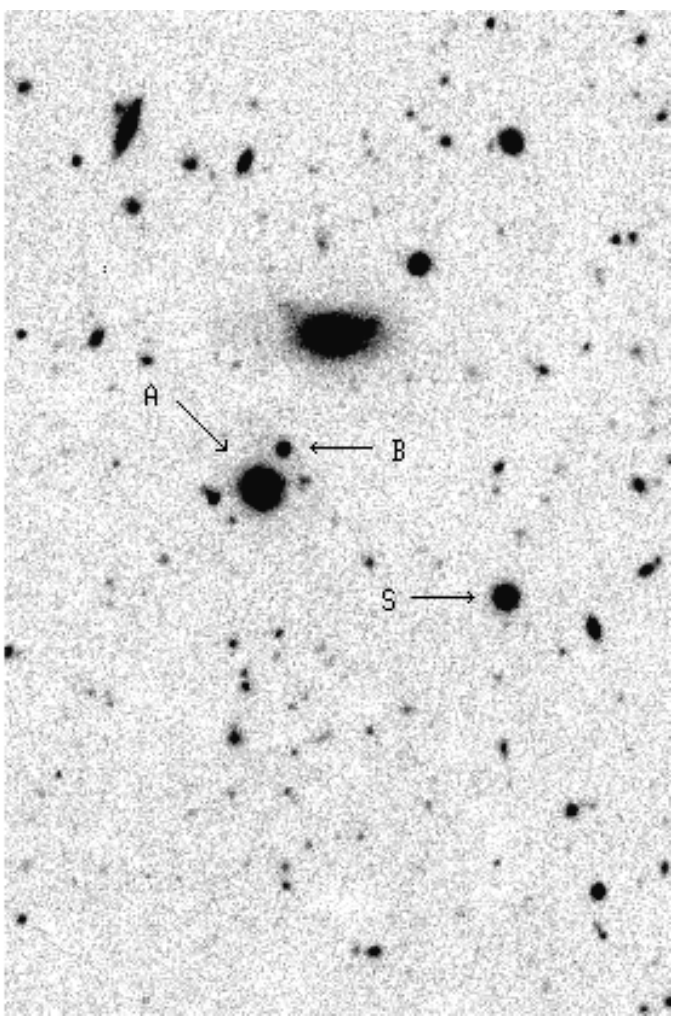

**Fig. 1.** Field of $1.7' \times 3.0'$ around UM 425. North-East is to the upper-left. This image is a combination of 29 R images used to obtain the light curves. The limiting R magnitude is about 24. The arrows show UM 425A, UM 425B, as well as the reference star (S), used to derive the photometry.

## 2. Observations

A number of CCD images of UM 425 have been obtained at ESO-La Silla from 1987 to 1994, mainly during the ESO Key-Programme on gravitational lenses (Surdej et al. 1989). The images were obtained with the 1.5-m Da-

**Table 1.** List of observations of UM 425 available from 1987 to 1994. 32 useful frames, out of this list of 53 images, were used. All the images used for the analysis are indicated by an asterisk, in the "Tel" column.

| Date | Tel | CCD num | pix (″) | Exp sec | Sky $e^-$ | Seeing (″) |
|---|---|---|---|---|---|---|
| 1987.32 | 2.2m | 5 | 0.351 | 60 | 805 | 1.1 |
| ″ | ″ * | ″ | ″ | 300 | 3925 | 1.3 |
| 1988.39 | 2.2m* | 5 | 0.351 | 300 | 6205 | 1.3 |
| ″ | ″* | ″ | ″ | 300 | 6275 | 1.2 |
| 1989.27 | 1.5m | 5 | 0.467 | 600 | 2520 | 1.4 |
| ″ | ″* | ″ | ″ | 300 | 635 | 1.4 |
| 1989.34 | 1.5m | 5 | 0.467 | (?) | 13400 | 2.4 |
| 1989.45 | 1.5m* | 5 | 0.467 | 180 | 10430 | 1.3 |
| ″ | ″ | ″ | ″ | 480 | 28380 | 1.4 |
| 1990.01 | 1.5m* | 11 | 0.232 | 600 | 11350 | 1.0 |
| ″ | ″* | ″ | ″ | 600 | 1888 | 1.0 |
| 1990.15 | 1.5m | 15 | 0.232 | 1200 | 3180 | 0.9 |
| ″ | ″* | ″ | ″ | 300 | 1935 | 0.9 |
| 1990.16 | 3.5m* | 17 | 0.153 | 300 | 688 | 0.8 |
| ″ | ″ | ″ | ″ | 500 | 1127 | 0.8 |
| 1990.21 | 2.2m | 11 | 0.175 | 600 | 1510 | 1.0 |
| ″ | ″* | ″ | ″ | 600 | 1525 | 1.0 |
| 1990.33 | 1.5m* | 15 | 0.232 | 900 | 23860 | 1.4 |
| ″ | ″* | ″ | ″ | 900 | 18270 | 1.4 |
| 1990.37 | 3.5m* | 17 | 0.153 | 240 | 745 | 1.2 |
| ″ | ″* | ″ | ″ | 240 | 715 | 0.9 |
| ″ | ″* | ″ | ″ | 120 | 325 | 0.7 |
| ″ | ″* | ″ | ″ | 120 | 325 | 0.7 |
| 1990.45 | 1.5m | 5 | 0.467 | 900 | 8760 | 0.8 |
| ″ | ″ | ″ | ″ | 300 | 2760 | 0.8 |
| 1990.47 | 1.5m | 5 | 0.467 | 900 | 11170 | 0.9 |
| ″ | ″ | ″ | ″ | 300 | 3480 | 0.9 |
| ″ | ″ | ″ | ″ | 120 | 1380 | 0.9 |
| 1990.48 | 1.5m | 5 | 0.467 | 900 | 19715 | 1.1 |
| ″ | ″ | ″ | ″ | 120 | 2640 | 1.2 |
| 1990.97 | 1.5m | 5 | 0.467 | 900 | 8710 | 1.2 |
| 1991.12 | 1.5m | 5 | 0.467 | 60 | 670 | 1.0 |
| ″ | ″* | ″ | ″ | 240 | 1810 | 1.0 |
| ″ | ″* | ″ | ″ | 240 | 1810 | 0.9 |
| ″ | ″* | ″ | ″ | 240 | 1790 | 0.9 |
| 1991.14 | 1.5m* | 15 | 0.232 | 1200 | 3140 | 1.0 |
| 1991.30 | 1.5m | 5 | 0.467 | 60 | 1105 | 1.0 |
| ″ | ″* | ″ | ″ | 480 | 5510 | 1.2 |
| ″ | ″* | ″ | ″ | 300 | 7790 | 1.1 |
| ″ | ″* | ″ | ″ | 600 | 11280 | 1.5 |
| 1991.36 | 3.5m | 24 | 0.316 | 120 | 5250 | 1.3 |
| ″ | ″* | ″ | ″ | 240 | 11050 | 1.3 |
| 1991.44 | 1.5m | 5 | 0.467 | 600 | 6300 | 1.1 |
| ″ | ″* | ″ | ″ | 300 | 3060 | 1.4 |
| 1992.08 | 1.5m | 7 | 0.340 | 60 | 950 | 1.0 |
| 1992.10 | 1.5m* | 13 | 0.232 | 600 | 295 | 1.2 |
| ″ | ″* | ″ | ″ | 600 | 285 | 1.3 |
| 1992.11 | 1.5m* | 13 | 0.232 | 1200 | 6310 | 1.4 |
| 1993.08 | 1.5m* | 28 | 0.371 | 600 | 2760 | 1.4 |
| ″ | ″* | ″ | ″ | 300 | 1440 | 1.3 |
| 1994.24 | 1.5m* | 28 | 0.371 | 180 | 14730 | 1.0 |
| 1994.35 | 3.5m | 36 | 0.268 | 900 | 8230 | 0.8 |
| ″ | ″* | ″ | ″ | 300 | 2775 | 0.8 |

Telescope (NTT). Table 1 gives the date of each observation, the telescope used and the number of the ESO CCD mounted on the instrument. The pixel size in arcseconds per pixel is also given, as well as the exposure time in seconds, the sky level in electrons and the seeing in arcseconds. Since component B is quite faint, a Bessel R filter has been used in order to maximize the efficiency of the observations.

During the seven years of the photometric monitoring, 53 frames were obtained. Most of the observations were achieved during the ESO Key-Programme on gravitational lenses and the frequency of the observations reached about one image per week during a few periods.

The main problem encountered in obtaining reliable data was mainly due to the large difference in brightness between the two components. So, the exposure time was chosen in order to have enough signal in the faint image, without saturating the bright one. The few saturated images were not retained for the analysis, nor were those for which image B was underexposed. The light curves of the A and B images have been obtained by comparison with the only useful reference star present on all frames (see Fig. 1).

## 3. Data reduction

Since the period of observation covers several years, and since it includes numerous telescope/detector configurations as well as variable weather conditions, the quality of the data is inhomogeneous. The same reduction procedure was nevertheless applied to all frames.

After bias subtraction and division by a flat-field, the sky level was estimated by fitting a low order polynomial surface through a number of "sky windows" selected interactively, and then subtracted. All the images with obvious defects after reduction, i.e. with imperfect flat-field correction or with artefacts produced by reflections of the moonlight on metallic parts of the telescope, were not included in the analysis. Some images with strong column offsets (CCD ESO#15) were not taken into account either. Finally, 32 frames were effectively useful for the photometry of the two images of UM 425.

In addition, a problem specific to CCD ESO #5 had to be corrected before any other treatment could be applied. The non-linearity of this CCD forced us to apply the following correction to all the frames taken with CCD #5, provided that the signal was higher than 4000 ADUs:

$$I_{corrected} = \frac{I_{raw}}{1.0382 - 9.5565 \cdot 10^{-6} \cdot I_{raw}}, \quad (1)$$

where $I_{corrected}$ is the intensity of a given pixel in the frame and $I_{raw}$ is the intensity of this pixel in the raw image (in ADUs). This formula was derived by Magain et

Although the separation between the two images is rather large ($\sim 6.5''$), it is not yet entirely sufficient to avoid mutual contaminations. Thus, aperture photometry does not give reliable results, and therefore a profile fitting method, developed by Remy (1995), must be preferred. This procedure allows to fit analytical or numerical functions, depending on a number of parameters, onto the image by minimizing the $\chi^2$.

The profiles fitted on the objects are two-dimensional Moffat profiles (Moffat 1969) with elliptical isophotes. The expression of one profile is given by the following formula:

$$\begin{aligned}I(x,y) =\ & P_1 \cdot [1 + (\frac{x-P_2}{P_4})^2 + (\frac{y-P_3}{P_5})^2 \\ & -2 \cdot P_6 \cdot (\frac{x-P_2}{P_4}) \cdot (\frac{y-P_3}{P_5})]^{-P_7},\end{aligned} \quad (2)$$

where (x,y) are the coordinates of the pixels, $P_1$ is the profile intensity, $P_2$ and $P_3$ are the coordinates of the center, $P_4$, $P_5$ and $P_6$ are the length, width and orientation of the elliptical isophotes of the 2-D Moffat profile. The value of exponent $P_7$ is fixed at 3.5, which generally gives the smallest residuals after subtraction of the synthetic profile from the image.

The fit is obtained in two steps:

1. We determine parameters $P_4$, $P_5$ and $P_6$ by fitting a single two-dimension profile on the reference star S (Fig. 1).
2. These parameters are used to fit simultaneously two Moffat profiles onto UM 425A and UM 425B, keeping the other 3 parameters free for each profile.

As the shape of the profile is the same for the reference star and for UM 425 (i.e. the same parameters $P_4$, $P_5$ and $P_6$), the photometry is obtained by comparing the values of parameter $P_1$ of the 3 fitted profiles. We also made the same kind of study with a numerical Point Spread Function constructed from the image of stars in the field. The results obtained are very similar to those obtained from the fit of 2-D Moffat profiles.

The uncertainties due to the fit and to the readout noise are quite negligible, at least in the faint component, in comparison to the uncertainties introduced by the photon noise. Thus, a good estimate of the relative $1\sigma$ error bars can be given, for this image, by:

$$\frac{\Delta F}{F} = \frac{\sqrt{F_{QSO} + F_{sky}}}{F_{QSO}}, \quad (3)$$

which is the photon noise in Poisson statistics. $F_{QSO}$ is the flux coming from the object and $F_{sky}$ is the flux from the sky, in an area equal to the size of the objects (Full Width at Zero Intensity) on the CCD. The fact that the Moffat

Function (PSF) can introduce a bias in the estimate of the uncertainties on the bright component. We have only considered here the uncertainties due to the photon noise.

## 4. Light curves - Discussion

The photometric data obtained as described in the previous section are presented in Fig. 2. From top to bottom are given, the light curve of UM 425B relative to UM 425A and the independent light curves of UM 425B and UM 425A relative to a reference star (see Fig. 1) in the field. The scatter in the data points shows that the estimated error bars are realistic for the faint component, but that they are underestimated for the brightest one, as expected. The relative light curve displays a strong peak in intensity. The fact that the variation is present in the light curve of B relative to A, as well as that of B relative to the reference star S, indicates that it is due to a variation of UM 425B itself.

On the other hand, we cannot exclude that the slow variation in the light curve of UM 425A (0.2 magnitude in 7 years) may be due to a variation of the reference star S, since the light curve of UM 425B seems to show the same smooth slope outside the strong peak. However, such a variation is not typical of stellar variability.

The strong peak present in the light curve of the B image can be interpreted in two ways. First, it could be due to an intrinsic variation of the quasar, in which case the fact that a similar peak is not observed in the light curve of the A image allows to set a lower limit on the time delay and, thus, an upper limit on the Hubble constant, if UM 425 is a gravitationally lensed system. The second interpretation is in terms of a microlensing event caused by a star in the deflecting galaxy. Both hypotheses will be examined in turn.

### 4.1. Intrinsic variation of the source

Assuming an intrinsic variation of the quasar luminosity, our observations would indicate a lower limit of 3 years on the time delay in this system, because the variation is not observed in UM 425A. A rough upper limit on the Hubble constant $H_0$ can be estimated by using a point mass model for the deflector. As the deflector is not detected (see next section) in the case of UM 425, the geometry of the system is not precisely known. However, the large angular separation between the components of UM 425, and their large difference in brightness, suggest that the deflector is very close (angle-wise) to UM 425B, at least if this deflector can be approximated by a point mass (Kassiola & Kovner 1992, Refsdal & Surdej 1994).

The difference between the spectrum of UM 425B and the scaled version of UM 425A leaves a residual, that Meylan & Djorgovski (1989) interpret as the spectrum of the deflector, the Hubble constant can be deduced from the time delay by the following formula (Refsdal & Surdej 1994):

$$H_0^{-1} = \frac{2(z_s - z_d)\Delta t}{z_s z_d \theta_{AB}(\theta_A + \theta_B)(2 - \epsilon)}. \tag{4}$$

where $z_s$ and $z_d$ are the redshifts of the source and of the deflector, respectively, and $\Delta t$ is the time delay. $\theta_{AB}$ is the angular separation between the images A and B of the quasar and $\theta_A$, $\theta_B$ are the angular separations between the deflector and each image. The value of $\epsilon$ depends on the nature of the adopted deflector. We thus have $\epsilon = 0$ for a point source and $\epsilon = 1$ for a SIS model. As we do not see the deflector, these two angles remain unknown but, as the deflector is expected to be very close to UM 425B, we can assume that $\theta_B \sim 0$ and that $\theta_A \sim \theta_{AB}$. Equation (4) then becomes:

$$H_0^{-1} \sim \frac{2(z_s - z_d)\Delta t}{z_s z_d \theta_{AB}^2 (2 - \epsilon)}. \tag{5}$$

With a lower limit of 3 years on $\Delta t$, we derive an upper limit for the Hubble constant consistent with any recent estimate of $H_0$ ($H_0 < 360\, km\, s^{-1}\, Mpc^{-1}$!).

Finally, had we assumed another lens model, such as an SIS model (Singular Isothermal Sphere), which is a reasonable representation of an elliptical galaxy, the upper limit on the Hubble constant would be two times smaller, which is again too high to give a useful constraint on $H_0$. The large angular separation between the components of UM 425 implies that the time delay expected for this system is quite long. Indeed, assuming a point mass model and that $50\, km\, s^{-1}\, Mpc^{-1} \lesssim H_0 \lesssim 100\, km\, s^{-1}\, Mpc^{-1}$, one gets $10\, years \lesssim \Delta t \lesssim 20\, years$. It is thus not surprising that our lower limit on the time delay (3 years) leads to a Hubble constant consistent with all recent estimate of $H_0$. In addition, an intrinsic variation of the source is expected to occur in UM 425A first and, then in UM 425B (Schneider et al. 1992, Refsdal & Surdej 1993). So, equation (5) suggests that a peak of intensity observed in the light curve of UM 425A, 10 to 20 years before 1990 would be in good agreement with the hypothesis of an intrinsic variation of the source. Old data, for example taken from Schmidt plates, could be useful in order to know if such an event ever occured in UM 425A.

### 4.2. Microlensing event

As indicated above, the hypothetical deflector is expected to be very "close" to UM 425B, hence increasing the probability that a star crosses the light path to component B of the quasar. Its apparent brightness could then increase

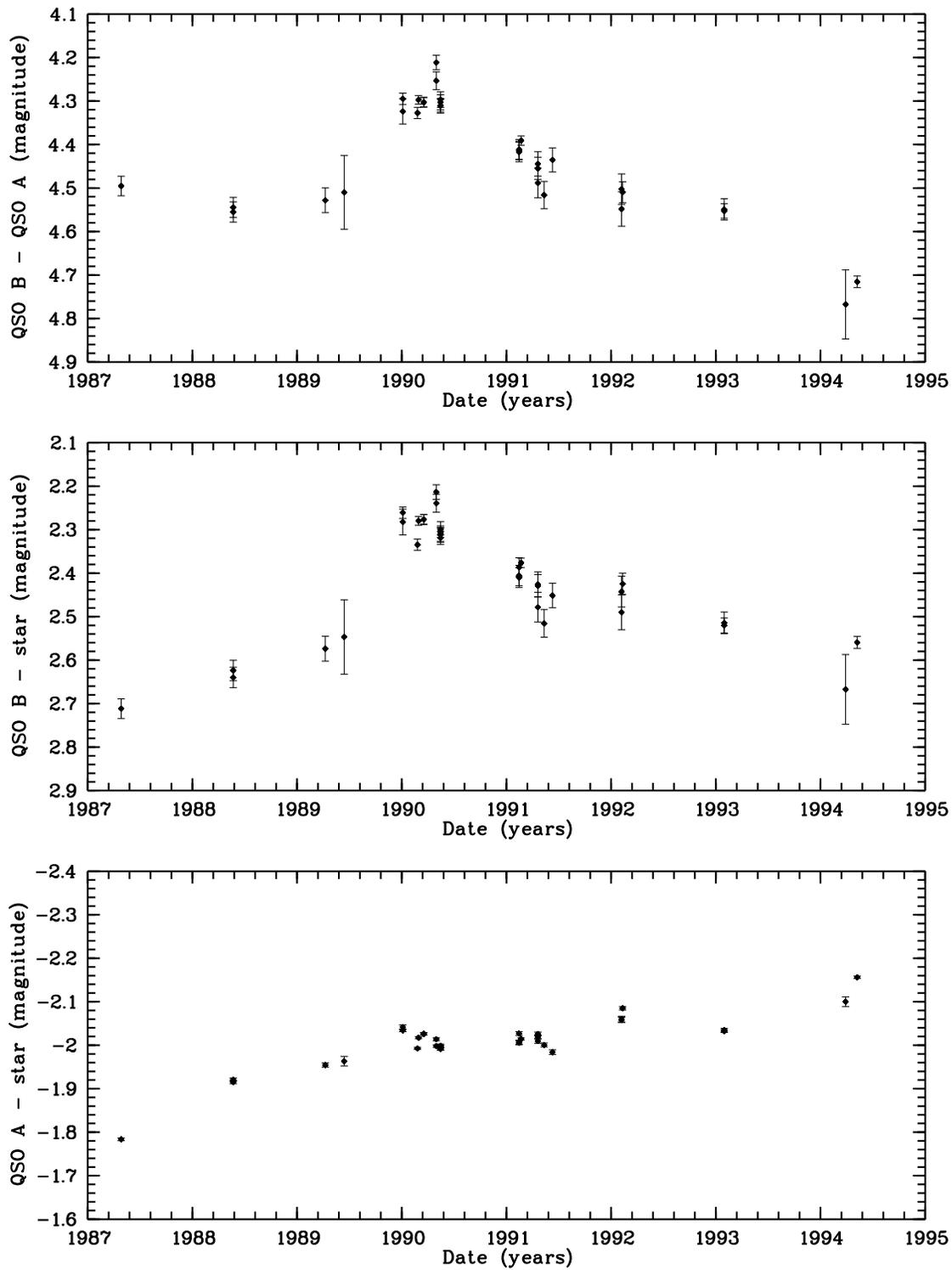

**Fig. 2.** From top to bottom: (top) light curve of UM 425 B relative to UM 425 A, from April 1987 to May 1994; (middle) light curve of UM 425 B, and (bottom) light curve of UM 425 A relative to the reference star S. The comparison between the curves of UM 425 A and UM 425 B clearly shows that the intensity peak in the relative light curve is real and is, in fact, only due to component B. One-sigma error bars are indicated (see text).

Assuming that the strong increase in the light curve of UM 425B is due to a microlensing "High Amplification Event", and assuming a typical transverse velocity of $V_t = 600\,km\,s^{-1}$ for the star in the deflecting galaxy (Schneider et al. 1992), an estimate of the size X of the emitting region in the quasar can be derived from the duration of the "peak", $t \sim 2$ years:

$$X \sim V_t \cdot t \cdot \frac{D_s}{D_d}, \qquad (6)$$

where $D_s$ is the distance from the source to the observer, and $D_d$ the distance between the observer and the deflector. The distances used are the "angular diameter distances" (Refsdal 1966). The peak of intensity lasts for 2 years (FWZI), thus, assuming that $q_0 = 0.5$ and $H_0 = 75\,km\,s^{-1}\,Mpc^{-1}$ we obtain a diameter of $\sim 2 \cdot 10^{-3}$pc for the amplified source, which is compatible with the size of an accretion disk, according to the standard models of AGNs (e.g. Blandford, Netzer & Woltjer 1990) and, thus, is in good agreement with the interpretation of the observed peak in terms of a microlensing "High Amplification Event".

However, the confirmation of a microlensing event would need further investigation, especially via spectroscopy, in order to detect variations in the line/continuum ratio in component B, if such an amplification occurs again.

The microlensing hypothesis can also be checked by analysing the light curve of UM 425 on a much longer time scale, by continuing the monitoring. If such peaks of intensity were observed only in the light curve of UM 425B, this would support the microlensing hypothesis (or the fact that UM 425 is not a lensed system).

## 5. Deep imaging of the field of UM 425

The frames obtained for the monitoring program were also co-added in order to try to detect the deflector or any faint object close to the quasar images. For that purpose, a weighted sum of these images was constructed, the weights being computed in order to optimize the detection of faint features. These adopted weights are, in fact, proportional to the ratio of the central intensity of a stellar image (the same star for all the images) to the standard deviation in the sky background. They thus correspond to the central Signal to Noise ratio in faint objects and roughly optimize their detection.

The final image is presented in Fig. 1. The equivalent exposure time is about one hour with a 4 meter class telescope. The faintest objects visible on this image have a magnitude of $m_R \sim 24$. This limiting magnitude is the magnitude of the faintests objects in the field, above a level of $3\sigma_{sky}$. Fig. 3 shows an enlargement of the central part of Fig. 1. The pixel size is $0\rlap{.}''175$ and the seeing

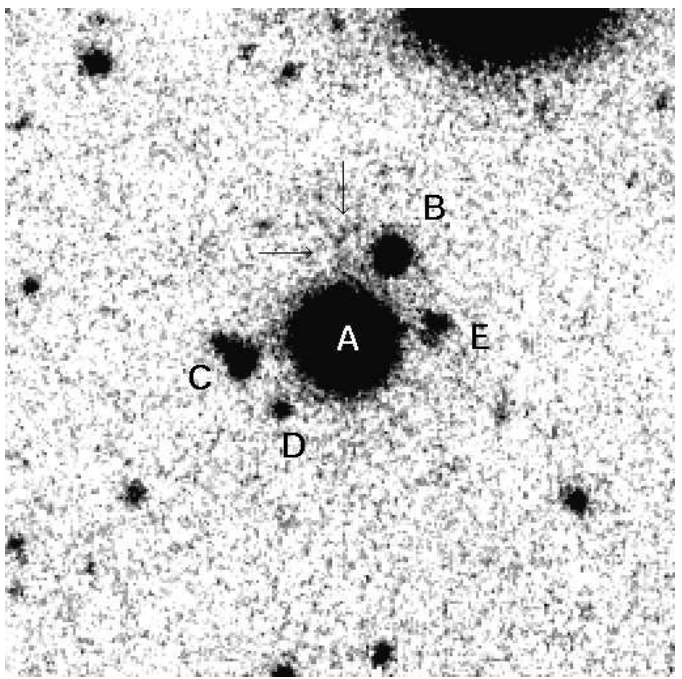

**Fig. 3.** Field of $45'' \times 45''$ around UM 425. North-East is at the upper-left, and UM 425 A is the brightest object in the center of the field. The pixel size is $0\rlap{.}''175$ and the seeing (FWHM) is $1''$. All the objects are labelled as indicated in the text. The arrows point to the faint feature, north of UM 425 A (see also Fig. 4).

(FWHM) is about $1''$.

At least four objects closer than $10''$ from UM 425A are detected. Their magnitudes are given in Table 2, but since the relative luminosities of the objects (at least for A and B) are different on the images used in our co-addition, these results are only rough estimates. The calibration for the determination of the magnitudes has been done by using the R magnitude of UM 425A given by Meylan & Djorgovski (1989), although this magnitude is affected by an error of a few percent. The relative positions were computed by fitting a two-dimensional gaussian profile to the objects.

Hes and Smette (1995) analyze a spectrum of UM 425C and show that this component is definitely not a quasar but a galaxy. To our knowledge, no spectrum has been obtained for UM 425D and its nature is still unknown. However, its fuzzy aspect seems to indicate that it is probably an intervening galaxy, which might play a role in the gravitational potential.

An additional diffuse object, UM 425E, is detected on our composite image. One can also notice a very faint feature, north of UM 425A (arrows on Fig. 3 and Fig. 4). Our data do not allow to ascertain the nature of this feature, which might also be an artefact on one or several image(s) in the sum.

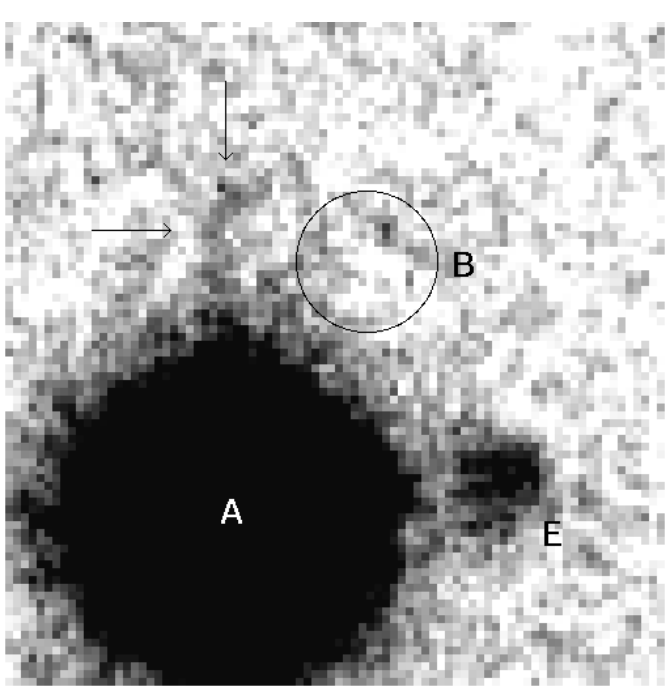

**Fig. 4.** Result of PSF subtraction performed on 29 individual images of UM 425. The residuals were co-added as exposed in the text. The field is $14'' \times 14''$ (North-East at the upper-left), with a pixel size of $0\rlap{.}''175$. The circle is centered at the position of UM 425B and its diameter corresponds to the FWZI of UM 425B ($3.2''$). No significant residuals can be seen, although the deflector is expected to be very close to UM 425B.

Taking into account only the images with a good sampling, we investigated PSF subtraction on UM 425A and B. Since the quasar's components are variable and that the individual PSFs are not axisymetric, the subtraction could not be performed on the co-added image because the profiles of the star and of the A and B components on the co-added final image are not identical anymore. We thus constructed the empirical PSFs, for each frame, using the only available star in field (star S in Fig. 1). Since UM 425A is much brighter than the star, the PSF obtained did not provide residuals of high quality after subtraction from this component. However, as explained earlier, the deflector responsible for the lensing effect is expected to be very close, angle-wise, to UM 425B. Thus, we subtracted the PSF from UM 425B and co-added the residuals obtained with the weighting system described previously. The result of such a process is shown in Fig 4. The final limiting magnitude is $m_R \sim 24$ ($3\sigma_{sky}$ above

**Table 2.** Relative positions of the objects around UM 425A and their R magnitude. The magnitudes of components C, D and E were derived by aperture photometry, because of their fuzzy aspect. Note that the uncertainty on the magnitude of UM 425B takes into account the fact that this component is highly variable.

| Object  | $\Delta x$ ($''$) | $\Delta y$ ($''$) | $m_R$          |
|---------|-------------------|-------------------|----------------|
| UM 425A | 0                 | 0                 | $15.7 \pm 0.2$ |
| UM 425B | $+3.2 \pm 0.1$    | $+5.7 \pm 0.1$    | $20.1 \pm 0.6$ |
| UM 425C | $-6.8 \pm 0.2$    | $-1.6 \pm 0.2$    | $21.3 \pm 0.2$ |
| UM 425D | $-4.0 \pm 0.2$    | $-4.7 \pm 0.2$    | $23.1 \pm 0.2$ |
| UM 425E | $+5.9 \pm 0.2$    | $+1.0 \pm 0.2$    | $22.4 \pm 0.2$ |

the sky level). The image has been enlarged, to show the details of the residuals.

Even if the subtraction is not perfect, we find no obvious deflector close to UM 425B, down to a R magnitude of 24. However, we hope to obtain soon deeper imaging, with data of homogeneous quality, with a better resolution and sampling, to draw a definite conclusion concerning the detection of the deflector and the nature of the feature North of UM 425A.

## 6. Conclusion

This 7 year monitoring allowed to investigate the light curves of components A and B of the gravitational lens candidate UM 425. The strong peak in intensity present in the relative light curve appears to be due to a variation of the faint component of the system.

There are two possible explanations for this variation. First it could be due to an intrinsic variation of the quasar, in which case a lower limit of 3 years on the time delay can be derived, which is still too low to constrain $H_0$. Second, it could be due to a microlensing event in the faint component of the system, giving an estimate of the size of the source of $\sim 2 \cdot 10^{-3}$pc. This size, fully compatible with the sizes expected for AGNs (the continuum region), seems to support the latter interpretation. However, the possibility of a microlensing effect has still to be confirmed, as well as the lensed nature of UM 425.

Our deep image does not allow the detection of the deflector nor of any arc or arclet, down to a limiting magnitude of $m_R \sim 24$. We do, however, detect one object, UM 425E, not observed by Meylan and Djorgovski, as well as a fuzzy extension, North of UM 425A, whose nature is doubtful. Deeper imaging and spectroscopic studies of this system have to be carried out in order to draw definite conclusions concerning the nature of the objects surrounding UM 425A.

Pursuing the monitoring of this gravitational lens candi-crolensing events, and for the determination of the Hubble constant $H_0$.

*Acknowledgements.* This work has been supported in part by the European Community HCM Network CHRX-CT92-0044. The research of the Liège team on gravitational lenses is supported in part by the Communauté Française and the Fonds National de la Recherche Scientifique of Belgium, and by contracts ARC 90/94-140 "Action de Recherche Concertée de la Communauté Française" (Belgium), SC 005 "Service Center and Research Networks" and GL/FOC/HST "PRODEX" both of the "Services Scientifiques, Techniques et Culturels" (Prime Ministers' office, Belgium).
A.S. receives financial support under grant no. 781-73-058 from the Netherlands Foundation for research in Astronomy (AS-TRON) which receives its funds from the Netherlands Organisation for Scientific Research (NWO).
E.V.D. is financially supported by the Japanese government (Monbusho scholarship).
It is a pleasure to thank Dr. J.P. Swings for carefully reading the manuscript, and for his help in improving the paper. The authors also thank P. Barthel who provided two additional images for the point of 1990.16 on the light curves.